%% file: main.tex
\documentclass[sigconf]{acmart}
\pdfoutput=1
\input{figs/qcircuit_common.tex}

\newcommand{\ignore}[1]{}
\usepackage{fancyhdr}
\usepackage[normalem]{ulem}
\usepackage{microtype}

\usepackage{pgfplots}
\pgfplotsset{compat=1.14}
\usetikzlibrary{patterns}
\usetikzlibrary{arrows}
\usepackage{xstring}

\usepackage{qcircuit}
\usepackage[boxed]{algorithm2e}
\usepackage{braket}
\usepackage{xcolor}
\usepackage{tabularx}
\usepackage{commath}
\usepackage{textcomp}

\usepackage{tikz}

\usetikzlibrary{automata, positioning, arrows, decorations.markings}
\tikzset{myptr/.style={decoration={markings,mark=at position 1 with %
    {\arrow[scale=2,>=stealth]{>}}},postaction={decorate}}}

\newcommand{\colorQubit}{blue}
\newcommand{\colorQubitBB}{gray}
\newcommand{\colorQutrit}{orange!70!red}
\newcommand{\colorLightness}{60}
\newcommand{\colorQubitLight}{\colorQubit!\colorLightness}

\newcommand{\colorQutritLight}{\colorQutrit!\colorLightness}
\definecolor{brown}{RGB}{77, 51, 0}

\title{Asymptotic Improvements to Quantum Circuits via Qutrits} 
\author{Pranav Gokhale}
\email{pranavgokhale@uchicago.edu}
\orcid{0000-0003-1946-4537}
\affiliation{
  \institution{University of Chicago}
}

\author{Jonathan M. Baker}
\email{jmbaker@uchicago.edu}
\affiliation{
  \institution{University of Chicago}
}

\author{Casey Duckering}
\email{cduck@uchicago.edu}
\affiliation{
  \institution{University of Chicago}
}

\author{Natalie C. Brown}
\email{natalie.c.brown@duke.edu}
\affiliation{
  \institution{Georgia Institute of Technology}
}

\author{Kenneth R. Brown}
\email{kenneth.r.brown@duke.edu}
\affiliation{
  \institution{Duke University}
}

\author{Frederic T. Chong}
\email{chong@cs.uchicago.edu}
\affiliation{
  \institution{University of Chicago}
}

\input{txt/0abstract.tex}



\begin{document}
\acmYear{2019}\copyrightyear{2019}
\setcopyright{acmlicensed}
\acmConference[ISCA '19]{ISCA '19: 46th International Symposium on Computer Architecture}{June 22--26, 2019}{PHOENIX, AZ, USA}
\acmBooktitle{ISCA '19: 46th International Symposium on Computer Architecture, June 22--26, 2019, PHOENIX, AZ, USA}
\acmPrice{15.00}
\acmDOI{10.1145/3307650.3322253}
\acmISBN{978-1-4503-6669-4/19/06}

\maketitle

\thispagestyle{firstpage}
\pagestyle{plain}

\input{txt/1introduction.tex}
\input{txt/2background.tex}
\input{txt/3related_work.tex}
\input{txt/4circuits.tex}
\input{txt/5applications_to_algorithms.tex}
\input{txt/6simulator.tex}
\input{txt/7noise_models.tex}
\input{txt/8results.tex}
\input{txt/9discussion.tex}

\section*{Acknowledgements}
We would like to thank Michel Devoret and Steven Girvin for suggesting that we investigate qutrits. We also acknowledge David Schuster for helpful discussion on superconducting qutrits. This work is funded in part by EPiQC, an NSF Expedition in Computing, under grants CCF-1730449/1832377, and in part by STAQ, under grant NSF Phy-1818914.

\appendix
\input{txt/detailed_noise_model.tex}

\bibliographystyle{ieeetr}
\bibliography{ref}

\end{document}

%% file: figs/qcircuit_common.tex
\usepackage[braket, qm]{qcircuit}
\usepackage{amsmath}
\usepackage{tikz}

\newcommand*\qcontrolcolor[2]{\push{\footnotesize \tikz[baseline=(char.base)]{
                              \node[shape=circle,draw,inner sep=0.6pt,color=#1] (char) {#2};}}
                              \qw}  
\newcommand*\qcontrol[1]{\qcontrolcolor{black}{#1}}
\newcommand*\onecontrol{\qcontrolcolor{red!80!black}{1}}
\newcommand*\twocontrol{\qcontrolcolor{blue!100!white}{2}}

%% file: txt/0abstract.tex
\begin{abstract}
Quantum computation is traditionally expressed in terms of quantum bits, or qubits. In this work, we instead consider three-level qu\textit{trits}. Past work with qutrits has demonstrated only constant factor improvements, owing to the $\log_2(3)$ binary-to-ternary compression factor. We present a novel technique using qutrits to achieve a logarithmic depth (runtime) decomposition of the Generalized Toffoli gate using no ancilla--a significant improvement over linear depth for the best qubit-only equivalent. Our circuit construction also features a 70x improvement in two-qudit gate count over the qubit-only equivalent decomposition. This results in circuit cost reductions for important algorithms like quantum neurons and Grover search. We develop an open-source circuit simulator for qutrits, along with realistic near-term noise models which account for the cost of operating qutrits. Simulation results for these noise models indicate over 90\% mean reliability (fidelity) for our circuit construction, versus under 30\% for the qubit-only baseline. These results suggest that qutrits offer a promising path towards scaling quantum computation.
\end{abstract}

%% file: txt/1introduction.tex
\section{Introduction}
Recent advances in both hardware and software for quantum computation have demonstrated significant progress towards practical outcomes. In the coming years, we expect quantum computing will have important applications in fields ranging from machine learning and optimization \cite{quantum_ml} to drug discovery \cite{quantum_medicine}. While early research efforts focused on longer-term systems employing full error correction to execute large instances of algorithms like Shor factoring \cite{Shor} and Grover search \cite{Grover}, recent work has focused on NISQ (Noisy Intermediate Scale Quantum) computation \cite{NISQ}. The NISQ regime considers near-term machines with just tens to hundreds of quantum bits (qubits) and moderate errors.

Given the severe constraints on quantum resources, it is critical to fully optimize the compilation of a quantum algorithm in order to have successful computation. Prior architectural research has explored techniques such as mapping, scheduling, and parallelism \cite{Mapping, Parallelism, IntelQuantumScheduling} to extend the amount of useful computation possible. In this work, we consider another technique: quantum trits (qutrits).

While quantum computation is typically expressed as a two-level binary abstraction of qubits, the underlying physics of quantum systems are not intrinsically binary. Whereas classical computers operate in binary states at the physical level (e.g. clipping above and below a threshold voltage), quantum computers have natural access to an infinite spectrum of discrete energy levels. In fact, hardware must actively suppress higher level states in order to achieve the two-level qubit approximation. Hence, using three-level qutrits is simply a choice of including an additional discrete energy level, albeit at the cost of more opportunities for error.

Prior work on qutrits (or more generally, d-level \textit{qudits}) identified only constant factor gains from extending beyond qubits. In general, this prior work \cite{Pavlidis} has emphasized the information compression advantages of qutrits. For example, $N$ qubits can be expressed as $\frac{N}{\log_2(3)}$ qutrits, which leads to $\log_2(3)\approx 1.6$-constant factor improvements in runtimes.

Our approach utilizes qutrits in a novel fashion, essentially using the third state as temporary storage, but at the cost of higher per-operation error rates. Under this treatment, the runtime (i.e. circuit depth or critical path) is \textit{asymptotically} faster, and the reliability of computations is also improved. Moreover, our approach only applies qutrit operations in an intermediary stage: the input and output are still qubits, which is important for initialization and measurement on real devices \cite{HesingA, HesingB}.

\begin{figure}[ht]
\centering
  \input{figs/graph_frontier.tikz}
  \caption{The frontier of what quantum hardware can execute is the yellow region adjacent to the 45\textdegree{} line. In this region, each machine qubit is a data qubit. Typical circuits rely on non-data ancilla qubits for workspace and therefore operate below the frontier.}
  \label{fig:frontier_of_QC}
\end{figure}
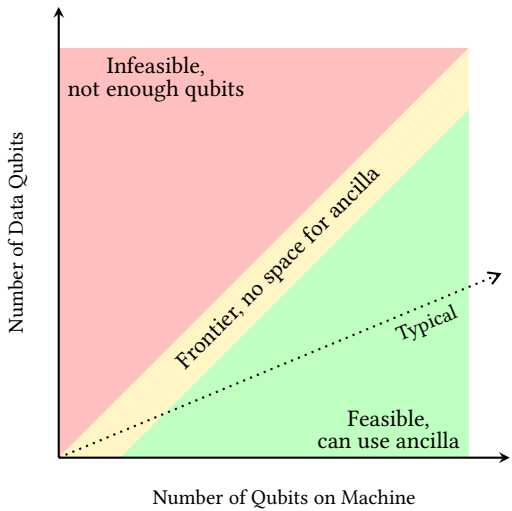

The net result of our work is to extend the frontier of what quantum computers can compute. In particular, the frontier is defined by the zone in which every machine qubit is a data qubit, for example a 100-qubit algorithm running on a 100-qubit machine. This is indicated by the yellow region in Figure~\ref{fig:frontier_of_QC}. In this frontier zone, we do not have room for non-data workspace qubits known as ancilla. The lack of ancilla in the frontier zone is a costly constraint that generally leads to inefficient circuits. For this reason, typical circuits instead operate below the frontier zone, with many machine qubits used as ancilla. Our work demonstrates that ancilla can be substituted with qutrits, enabling us to operate efficiently within the ancilla-free frontier zone.

We highlight the three primary contributions of our work:
\begin{enumerate}
    \item A circuit construction based on qutrits that leads to asymptotically faster circuits ($633N \rightarrow 38 \log_2 N$) than equivalent qubit-only constructions. We also reduce total gate counts from $397N$ to $6N$.
    \item An open-source simulator, based on Google's Cirq \cite{Cirq}, which supports realistic noise simulation for qutrit (and qudit) circuits.
    \item Simulation results, under realistic noise models, which demonstrate our circuit construction outperforms equivalent qubit circuits in terms of error. For our benchmarked circuits, our reliability advantage ranges from 2x for trapped ion noise models up to more than 10,000x for superconducting noise models. For completeness, we also benchmark our circuit against a qubit-only construction augmented by an ancilla and find our construction is still more reliable.
\end{enumerate}

The rest of this paper is organized as follows: Section~\ref{sec:background} presents relevant background about quantum computation and Section~\ref{sec:related_work} outlines related prior work that we benchmark our work against. Section~\ref{sec:circuit constructions} demonstrates our key circuit construction, and Section~\ref{sec:application to algorithms} surveys applications of this construction toward important quantum algorithms. Section \ref{sec:simulator} introduces our open-source qudit circuit simulator. Section~\ref{sec:error modeling} explains our noise modeling methodology (with full details in Appendix~\ref{app:detailed_noise_models}), and Section~\ref{sec:results} presents simulation results for these noise models. Finally, we discuss our results at a higher level in Section~\ref{sec:discussion}.

%% file: figs/graph_frontier.tikz
\begin{tikzpicture}
\pgfplotsset{every axis/.append style={thick}}
\pgfplotsset{every tick label/.append style={font=\small}}
\pgfplotsset{every axis label/.append style={font=\small}}

\definecolor{colorFrontierInfeasible}{rgb}{1.0,0.5,0.5}
\definecolor{colorFrontierNoAncilla}{RGB}{255,245,197}
\definecolor{colorFrontierFeasible}{rgb}{0.5,1.0,0.5}

\begin{axis}[
    xlabel=Number of Qubits on Machine,
    ylabel=Number of Data Qubits,
    width=250pt,
    xmin=0, xmax=110,
    ymin=0, ymax=110,
    axis equal image,
    axis on top,
    ,
    xticklabels={,,},
    yticklabels={,,},
    axis line style={draw=black},
    axis lines=left,
    tick style={draw=none},
    clip=false,
]

\addplot[draw=none,fill=colorFrontierInfeasible,fill opacity=0.5] coordinates {
    (0, 0)
    (0, 100)
    (100, 100)
};
\addplot[draw=none,fill=colorFrontierNoAncilla,fill opacity=1] coordinates {
    (0, 0)
    (15, 0)
    (100, 85)
    (100, 100)
};
\addplot[draw=none,fill=colorFrontierFeasible,fill opacity=0.5] coordinates {
    (15, 0)
    (100, 0)
    (100, 85)
};

\addplot[draw=none] coordinates {
    (0, 100)
}
node [anchor=north west, rotate=0, xshift=0em] {$\substack{\mbox{Infeasible,}\\\mbox{not enough qubits}}$};

\addplot[draw=none] coordinates {
    (100, 0)
}
node [anchor=south east, rotate=0, xshift=0em] {$\substack{\mbox{Feasible,}\\\mbox{can use ancilla}}$};

\addplot[draw=none] coordinates {
    (53.75, 46.25)
}
node [anchor=center, rotate=45, xshift=0em] {Frontier, no space for ancilla};


\draw[arrows={-angle 60}][dotted] (axis cs:0,0) -- (axis cs:108,45);
\addplot[draw=none] coordinates {
    (108, 45)
}
node [anchor=north east, rotate=22.62, xshift=-2em] {\small Typical};

\end{axis}
\end{tikzpicture}

%% file: txt/2background.tex

\section{Background} \label{sec:background}

A qubit is the fundamental unit of quantum computation. Compared to their classical counterparts which take values of either 0 and 1, qubits may exist in a superposition of the two states. We designate these two basis states as $\ket{0}$ and $\ket{1}$ and can represent any qubit as $\ket{\psi} = \alpha\ket{0} + \beta\ket{1}$ with $\|\alpha\|^2 + \|\beta\|^2 = 1$. $\|\alpha\|^2$ and $\|\beta\|^2$ correspond to the probabilities of measuring $\ket{0}$ and $\ket{1}$ respectively.

Quantum states can be acted on by quantum gates which (a) preserve valid probability distributions that sum to 1 and (b) guarantee reversibility. For example, the X gate transforms a state $\ket{\psi} = \alpha\ket{0} + \beta\ket{1}$ to $X\ket{\psi} = \beta\ket{0} + \alpha\ket{1}$. The X gate is also an example of a classical reversible operation, equivalent to the NOT operation. In quantum computation, we have a single irreversible operation called measurement that transforms a quantum state into one of the two basis states with a given probability based on $\alpha$ and $\beta$. 

In order to interact different qubits, two-qubit operations are used. The CNOT gate appears both in classical reversible computation and in quantum computation. It has a control qubit and a target qubit. When the control qubit is in the $\ket{1}$ state, the CNOT performs a NOT operation on the target. The CNOT gate serves a special role in quantum computation, allowing quantum states to become entangled so that a pair of qubits cannot be described as two individual qubit states. Any operation may be conditioned on one or more controls. 

\begin{figure}[h]
		\centering
		\input{figs/toffoli_gate.qcircuit}
		\caption{Reversible AND circuit using a single ancilla bit. The inputs are on the left, and time flows rightward to the outputs. This AND gate is implemented using a Toffoli (CCNOT) gate with inputs $q_0$, $q_1$ and a single ancilla initialized to 0. At the end of the circuit, $q_0$ and $q_1$ are preserved, and the ancilla bit is set to 1 if and only if both other inputs are 1.}
		\label{fig:reversible_AND}
	\end{figure}
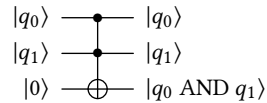

Many classical operations, such as AND and OR gates, are irreversible and therefore cannot directly be executed as quantum gates. For example, consider the output of 1 from an OR gate with two inputs. With only this information about the output, the value of the inputs cannot be uniquely determined. These operations can be made reversible by the addition of extra, temporary workspace bits initialized to 0. Using a single additional ancilla, the AND operation can be computed reversibly as in Figure \ref{fig:reversible_AND}.

Physical systems in classical hardware are typically binary. However, in common quantum hardware, such as in superconducting and trapped ion computers, there is an infinite spectrum of discrete energy levels. The qubit abstraction is an artificial approximation achieved by suppressing all but the lowest two energy levels. Instead, the hardware may be configured to manipulate the lowest three energy levels by operating on qutrits. In general, such a computer could be configured to operate on any number of $d$ levels, except as $d$ increases the number of opportunities for error, termed error channels, increases. Here, we focus on $d = 3$ with which we achieve the desired improvements to the Generalized Toffoli gate. 

In a three level system, we consider the computational basis states $\ket{0}$, $\ket{1}$, and $\ket{2}$ for qutrits. A qutrit state $\ket{\psi}$ may be represented analogously to a qubit as $\ket{\psi} = \alpha\ket{0} + \beta\ket{1} + \gamma\ket{2}$, where $\norm{\alpha}^2 + \norm{\beta}^2 + \norm{\gamma}^2 = 1$. Qutrits are manipulated in a similar manner to qubits; however, there are additional gates which may be performed on qutrits.

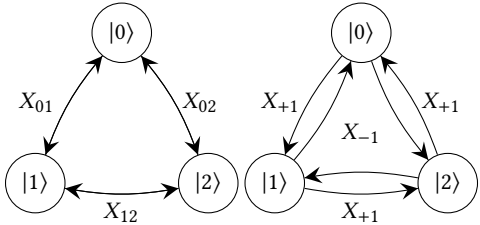
\begin{figure}[ht]
    \centering
        \begin{tikzpicture}[scale=1, every node/.style={scale=1}]
            \node[state] at (0,0) (0) {$\ket{0}$};
            \node[state] at (-1.15, -2) (1) {$\ket{1}$};
            \node[state] at (1.15, -2) (2) {$\ket{2}$};
            \draw   (0) edge[myptr, bend right=10, above, left=1] node{$X_{01}~$} (1)
                    (1) edge[myptr, bend left=10, above, right=1] (0)
                    (1) edge[myptr, bend right=10, below] node{$X_{12}$} (2)
                    (2) edge[myptr, bend left=10, above]  (1)
                    (0) edge[myptr, bend left=10, below, right=1] node{$X_{02}$} (2)
                    (2) edge[myptr, bend right=10, below, right=1] (0);
        \end{tikzpicture}~~~~~~~~
        \begin{tikzpicture}[scale=1, every node/.style={scale=1}]
            \node[state] at (0,0) (0) {$\ket{0}$};
            \node[state] at (-1.15, -2) (1) {$\ket{1}$};
            \node[state] at (1.15, -2) (2) {$\ket{2}$};
            \node at (0,-1.333) {$X_{-1}$};
            \draw   (0) edge[myptr, above, bend right=10, left=1] node{$X_{+1}$} (1)
                    (1) edge[myptr, above, bend right=10, right=1] (0)
                    (1) edge[myptr, below, bend right=10] node{$X_{+1}$} (2)
                    (2) edge[myptr, above, bend right=10] (1)
                    (0) edge[myptr, below, bend right=10, left=1] (2)
                    (2) edge[myptr, above, bend right=10, right=1] node{$X_{+1}$} (0);
        \end{tikzpicture}
    \caption{The five nontrivial permutations on the basis elements for a qutrit. (Left) Each operation here switches two basis elements while leaving the third unchanged. These operations are self-inverses. (Right) These two operations permute the three basis elements by performing a $+1\mod{3}$ and $-1\mod{3}$ operation. They are each other's inverses.}
    \label{fig:qutrit_xops}
\end{figure}

For instance, in quantum binary logic, there is only a single X gate. In ternary, there are three X gates denoted $X_{01}$, $X_{02}$, and $X_{12}$. Each of these $X_{ij}$ for $i \neq j$ can be viewed as swapping $\ket{i}$ with $\ket{j}$ and leaving the third basis element unchanged. For example, for a qutrit $\ket{\psi} = \alpha\ket{0} + \beta\ket{1} + \gamma\ket{2}$, applying $X_{02}$ produces $X_{02}\ket{\psi} = \gamma\ket{0} + \beta\ket{1} + \alpha\ket{2}$. Each of these operations' actions can be found in the left state diagram in Figure $\ref{fig:qutrit_xops}$.

There are two additional non-trivial operations on a single trit. They are the $+1$ and $-1$ (sometimes referred to as a $+2$) operations (with $+$ meaning addition modulo 3). These operations can be written as $X_{01}X_{12}$ and $X_{12}X_{01}$, respectively; however, for simplicity, we will refer to them as $X_{+1}$ and $X_{-1}$ operations. A summary of these gates' actions can be found in the right state diagram in Figure \ref{fig:qutrit_xops}.

Other, non-classical, operations may be performed on a single qutrit. For example, the Hadamard gate \cite{Nielsen} can be extended to work on qutrits in a similar fashion as the X gate was extended. In fact, all single qubit gates, like rotations, may be extended to operate on qutrits. In order to distinguish qubit and qutrit gates, all qutrit gates will appear with an appropriate subscript.

Just as single qubit gates have qutrit analogs, the same holds for two qutrit gates. For example, consider the CNOT operation, where an X gate is performed conditioned on the control being in the $\ket{1}$ state. For qutrits, any of the X gates presented above may be performed, conditioned on the control being in any of the three possible basis states. Just as qubit gates are extended to take multiple controls, qutrit gates are extended similarly. The set of single qutrit gates, augmented by any entangling two-qutrit gate, is sufficient for universality in ternary quantum computation \cite{brylinski2002universal}.

One question concerning the feasibility of using higher states beyond the standard two is whether these gates can be implemented and perform the desired manipulations. Qutrit gates have been successfully implemented \cite{Di, MS2000, Klimov2003} indicating it is possible to consider higher level systems apart from qubit only systems.

In order to evaluate a decomposition of a quantum circuit, we consider quantum circuit costs. The space cost of a circuit, i.e. the number of qubits (or qutrits), is referred to as circuit \textit{width}. Requiring ancilla increases the circuit width and therefore the space cost of a circuit. The time cost for a circuit is the \textit{depth} of a circuit. The depth is given as the length of the critical path (in terms of gates) from input to output.

%% file: figs/toffoli_gate.qcircuit

%

\[
\Qcircuit @R=1em @C=1em @!=0.5em {
\lstick{\ket{q_{0}}} & \control \qw & \rstick{\ket{q_{0}}} \qw & \\
\lstick{\ket{q_{1}}} & \control \qw\qwx  & \rstick{\ket{q_{1}}} \qw & \\
\lstick{\ket{0}} & \targ\qwx  & \rstick{\ket{q_0 \mbox{ AND } q_1}} \qw & \\
}
\]

%% file: txt/3related_work.tex
\section{Prior Work} \label{sec:related_work}

\subsection{Qudits}
Qutrits, and more generally qudits, have been been studied in past work both experimentally and theoretically. Experimentally, $d$ as large as 10 has been achieved (including with two-qudit operations) \cite{Kues}, and $d=3$ qutrits are commonly used internally in many quantum systems \cite{Zinner2018, Wallraff2018}.

However, in past work, qudits have conferred only an information compression advantage. For example, $N$ qubits can be compressed to $\frac{N}{\log_2(d)}$ qudits, giving only a constant-factor advantage \cite{Pavlidis} at the cost of greater errors from operating qudits instead of qubits. Under the assumption of linear cost scaling with respect to $d$, it has been demonstrated that $d=3$ is optimal \cite{Greentree, Khan}, although as we show in Section~\ref{sec:error modeling} the cost is generally superlinear in $d$.

The information compression advantage of qudits has been applied specifically to Grover's search algorithm \cite{YaleFan, Li, Wang, Ivanov} and to Shor's factoring algorithm \cite{Bocharov}. Ultimately, the tradeoff between information compression and higher per-qudit errors has not been favorable in past work. As such, the past research towards building practical quantum computers has focused on qubits.

Our work introduces qutrit-based circuits which are \textit{asymptotically} better than equivalent qubit-only circuits. Unlike prior work, we demonstrate a compelling advantage in both runtime and reliability, thus justifying the use of qutrits.

\subsection{Generalized Toffoli Gate}
We focus on the Generalized Toffoli gate, which simply adds more controls to the Toffoli circuit in Figure~\ref{fig:reversible_AND}. The Generalized Toffoli gate is an important primitive used across a wide range of quantum algorithms, and it has been the focus of extensive past optimization work. Table~\ref{tab:n_controlled} compares past circuit constructions for the Generalized Toffoli gate to our construction, which is presented in full in Section~\ref{subsec:generalized_toffoli_construction}.

\begin{table*}[]
\begin{tabular}{l|l|l|l|l|l|l}
  & \textbf{This Work} & Gidney \cite{GidneyBlogPost} & He \cite{He} & Barenco \cite{Barenco} & Wang \cite{Wang} & Lanyon \cite{Lanyon}, Ralph \cite{Ralph} \\ \hline
 Depth & $\log{N}$ & $N$ & $\log{N}$ & $N^2$ & $N$ & $N$ \rule{0pt}{2.6ex}
 \\ Ancilla & 0 & 0 & $N$ & 0 & 0 & 0
 \\ Qudit Types & Controls are qutrits & Qubits & Qubits & Qubits & Controls are qutrits & Target is $d=N$-level qudit
  \\ Constants & Small & Large & Small & Small & Small & Small
\end{tabular}
\caption{Asymptotic comparison of $N$-controlled gate decompositions. The total gate count for all circuits scales linearly (except for Barenco \cite{Barenco}, which scales quadratically). Our construction uses qutrits to achieve logarithmic depth without ancilla. We benchmark our circuit construction against Gidney \cite{GidneyBlogPost}, which is the asymptotically best ancilla-free qubit circuit.}
\label{tab:n_controlled}
\end{table*}

Among prior work, the Gidney \cite{GidneyBlogPost}, He \cite{He}, and Barenco \cite{Barenco} designs are all qubit-only. The three circuits have varying tradeoffs. While Gidney and Barenco operate at the ancilla-free frontier, they have large circuit depths: linear with a large constant for Gidney and quadratic for Barenco. The Gidney design also requires rotation gates for very small angles, which poses an experimental challenge. While the He circuit achieves logarithmic depth, it requires an ancilla for each data qubit, effectively halving the effective potential of any given quantum hardware. Nonetheless, in practice, most circuit implementations use these linear-ancilla constructions due to their small depths and gate counts.

As in our approach, circuit constructions from Lanyon \cite{Lanyon}, Ralph \cite{Ralph}, and Wang \cite{Wang} have attempted to improve the ancilla-free Generalized Toffoli gate by using qudits. Both the Lanyon \cite{Lanyon} and Ralph \cite{Ralph} constructions, which have been demonstrated experimentally, achieve linear circuit depths by operating the target as a $d=N$-level qudit. Wang \cite{Wang} also achieves a linear circuit depth but by operating each control as a qutrit.

Our circuit construction, presented in Section~\ref{subsec:generalized_toffoli_construction}, has similar structure to the He design, which can be represented as a binary tree of gates. However, instead of storing temporary results with a linear number of ancilla qubits, our circuit temporarily stores information directly in the qutrit $\ket{2}$ state of the controls. Thus, no ancilla are needed.

In our simulations, we benchmark our circuit construction against the Gidney construction \cite{GidneyBlogPost} because it is the asymptotically best qubit circuit in the ancilla-free frontier zone. We label these two benchmarks as \textcolor{brown}{QUTRIT} and \textcolor{brown}{QUBIT}. The \textcolor{brown}{QUBIT} circuit handles the lack of ancilla by using \textit{dirty} ancilla, which unlike \textit{clean} (initialized to $\ket{0}$) ancilla, can have an unknown initial state. Dirty ancilla can therefore be bootstrapped internally from a quantum circuit. However, this technique requires a large number of Toffoli gates which makes the decomposition particularly expensive in gate count.

Augmenting the base Gidney construction with a single ancilla\footnote{This ancilla can also also be dirty.} does reduce the constants for the decomposition significantly, although the asymptotic depth and gate counts are maintained. For completeness, we also benchmark our circuit against this augmented construction,  \textcolor{brown}{QUBIT+ANCILLA}. However, the augmented circuit does not operate at the ancilla-free frontier, and it conflicts with parallelism, as discussed in Section~\ref{sec:discussion}.

%% file: txt/4circuits.tex
\section{Circuit Construction} \label{sec:circuit constructions}

In order for quantum circuits to be executable on hardware, they are typically decomposed into single- and two- qudit gates. Performing efficient low depth and low gate count decompositions is important in both the NISQ regime and beyond. Our circuits assume all-to-all connectivity--we discuss this assumption in Section~\ref{sec:discussion}.

\subsection{Key Intuition}

To develop intuition for our technique, we first present a Toffoli gate decomposition which lays the foundation for our generalization to multiple controls. In each of the following constructions, all inputs and outputs are qubits, but we may occupy the $\ket{2}$ state temporarily during computation. Maintaining binary input and output allows these circuit constructions to be inserted into any preexisting qubit-only circuits.

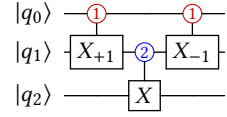
\begin{figure}
\[
    \Qcircuit @R=0.5em @C=0.25em {
    \lstick{\ket{q_{0}}} & \onecontrol & \qw  & \onecontrol & \qw & \\
    \lstick{\ket{q_{1}}} & \gate{X_{+1}}\qwx  & \twocontrol & \gate{X_{-1}}\qwx  &  \qw & \\
    \lstick{\ket{q_{2}}} & \qw  & \gate{X} \qwx  & \qw  &  \qw & \\
    }
\]
\caption{A Toffoli decomposition via qutrits. Each input and output is a qubit. The red controls activate on $\ket{1}$ and the blue controls activate on $\ket{2}$. The first gate temporarily elevates $q_1$ to $\ket{2}$ if both $q_0$ and $q_1$ were $\ket{1}$. We then perform the X operation only if $q_1$ is $\ket{2}$. The final gate restores $q_0$ and $q_1$ to their original state.}

\label{fig:toffoli_decomposition}
\end{figure}

In Figure \ref{fig:toffoli_decomposition}, a Toffoli decomposition using qutrits is given. A similar construction for the Toffoli gate is known from past work \cite{Lanyon, Ralph}. The goal is to perform an X operation on the last (target) input qubit $q_2$ if and only if the two control qubits, $q_0$ and $q_1$, are both $\ket{1}$. First a $\ket{1}$-controlled $X_{+1}$ is performed on $q_0$ and $q_1$. This elevates $q_1$ to $\ket{2}$ iff $q_0$ and $q_1$ were both $\ket{1}$. Then a $\ket{2}$-controlled $X$ gate is applied to $q_2$. Therefore, $X$ is performed only when both $q_0$ and $q_1$ were $\ket{1}$, as desired. The controls are restored to their original states by a $\ket{1}$-controlled $X_{-1}$ gate, which undoes the effect of the first gate. The key intuition in this decomposition is that the qutrit $\ket{2}$ state can be used instead of ancilla to store temporary information.

\subsection{Generalized Toffoli Gate}
\label{subsec:generalized_toffoli_construction}

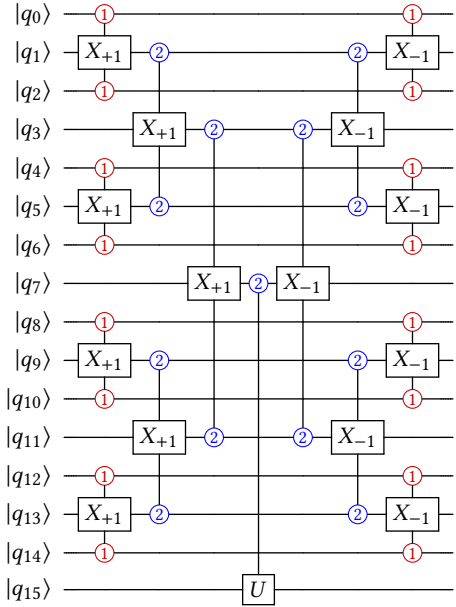
\begin{figure}[ht]
  \input{figs/BTBCnU.qcircuit}
  \caption{Our circuit decomposition for the Generalized Toffoli gate is shown for 15 controls and 1 target. The inputs and outputs are both qubits, but we allow occupation of the $\ket{2}$ qutrit state in between. The circuit has a tree structure and maintains the property that the root of each subtree can only be elevated to $\ket{2}$ if all of its control leaves were $\ket{1}$. Thus, the $U$ gate is only executed if all controls are $\ket{1}$. The right half of the circuit performs uncomputation to restore the controls to their original state. This construction applies more generally to any multiply-controlled $U$ gate. Note that the three-input gates are decomposed into 6 two-input and 7 single-input gates in our actual simulation, as based on the decomposition in \cite{Di}.}
  \label{fig:btb_cnu}
\end{figure}

We now present our circuit decomposition for the Generalized Toffoli gate in Figure~\ref{fig:btb_cnu}. The decomposition is expressed in terms of three-qutrit gates (two controls, one target) instead of single- and two- qutrit gates, because the circuit can be understood purely classically at this granularity. In actual implementation and in our simulation, we used a decomposition \cite{Di} that requires 6 two-qutrit and 7 single-qutrit physically implementable quantum gates.

Our circuit decomposition is most intuitively understood by treating the left half of the the circuit as a tree. The desired property is that the root of the tree, $q_7$, is $\ket{2}$ if and only if each of the 15 controls was originally in the $\ket{1}$ state. To verify this property, we observe the root $q_7$ can only become $\ket{2}$ iff $q_7$ was originally $\ket{1}$ and $q_3$ and $q_{11}$ were both previously $\ket{2}$. At the next level of the tree, we see $q_3$ could have only been $\ket{2}$ if $q_3$ was originally $\ket{1}$ and both $q_1$ and $q_5$ were previously $\ket{2}$, and similarly for the other triplets. At the bottom level of the tree, the triplets are controlled on the $\ket{1}$ state, which are only activated when the even-index controls are all $\ket{1}$. Thus, if any of the controls were not $\ket{1}$, the $\ket{2}$ states would fail to propagate to the root of the tree. The right half of the circuit performs \textit{uncomputation} to restore the controls to their original state.

After each subsequent level of the tree structure, the number of qubits under consideration is reduced by a factor of $\sim2$. Thus, the circuit depth is logarithmic in $N$. Moreover, each qutrit is operated on by a constant number of gates, so the total number of gates is linear in $N$.

Our circuit decomposition still works in a straightforward fashion when the control type of the top qubit, $q_0$, activates on $\ket{2}$ or $\ket{0}$ instead of activating on $\ket{1}$. These two constructions are necessary for the Incrementer circuit in \ref{sec:incrementer}.

We verified our circuits, both formally and via simulation. Our verification scripts are available on our GitHub \cite{QutritsGithub}.

%% file: figs/BTBCnU.qcircuit

%

\[
\Qcircuit @R=0.5em @C=0.1em {
\lstick{\ket{q_{0}}} & \qw & \qw & \qw & \qw & \qw & \onecontrol & \qw  & \qw  & \qw  & \qw  & \qw  & \onecontrol & \qw & \qw & \qw & \qw & \qw & \qw \\
\lstick{\ket{q_{1}}} & \qw & \qw & \qw & \qw & \qw & \gate{X_{+1}}\qwx  & \twocontrol & \qw  & \qw  & \qw  & \twocontrol & \gate{X_{-1}}\qwx  & \qw & \qw & \qw & \qw & \qw & \qw \\
\lstick{\ket{q_{2}}} & \qw & \qw & \qw & \qw & \qw & \onecontrol\qwx  & \qw \qwx  & \qw  & \qw  & \qw  & \qw \qwx  & \onecontrol\qwx  & \qw & \qw & \qw & \qw & \qw & \qw \\
\lstick{\ket{q_{3}}} & \qw & \qw & \qw & \qw & \qw & \qw  & \gate{X_{+1}}\qwx  & \twocontrol & \qw  & \twocontrol & \gate{X_{-1}}\qwx  & \qw  & \qw & \qw & \qw & \qw & \qw & \qw \\
\lstick{\ket{q_{4}}} & \qw & \qw & \qw & \qw & \qw & \onecontrol & \qw \qwx  & \qw \qwx  & \qw  & \qw \qwx  & \qw \qwx  & \onecontrol & \qw & \qw & \qw & \qw & \qw & \qw \\
\lstick{\ket{q_{5}}} & \qw & \qw & \qw & \qw & \qw & \gate{X_{+1}}\qwx  & \twocontrol\qwx  & \qw \qwx  & \qw  & \qw \qwx  & \twocontrol\qwx  & \gate{X_{-1}}\qwx  & \qw & \qw & \qw & \qw & \qw & \qw \\
\lstick{\ket{q_{6}}} & \qw & \qw & \qw & \qw & \qw & \onecontrol\qwx  & \qw  & \qw \qwx  & \qw  & \qw \qwx  & \qw  & \onecontrol\qwx  & \qw & \qw & \qw & \qw & \qw & \qw \\
\lstick{\ket{q_{7}}} & \qw & \qw & \qw & \qw & \qw & \qw  & \qw  & \gate{X_{+1}}\qwx  & \twocontrol & \gate{X_{-1}}\qwx  & \qw  & \qw  & \qw & \qw & \qw & \qw & \qw & \qw \\
\lstick{\ket{q_{8}}} & \qw & \qw & \qw & \qw & \qw & \onecontrol & \qw  & \qw \qwx  & \qw \qwx  & \qw \qwx  & \qw  & \onecontrol & \qw & \qw & \qw & \qw & \qw & \qw \\
\lstick{\ket{q_{9}}} & \qw & \qw & \qw & \qw & \qw & \gate{X_{+1}}\qwx  & \twocontrol & \qw \qwx  & \qw \qwx  & \qw \qwx  & \twocontrol & \gate{X_{-1}}\qwx  & \qw & \qw & \qw & \qw & \qw & \qw \\
\lstick{\ket{q_{10}}} & \qw & \qw & \qw & \qw & \qw & \onecontrol\qwx  & \qw \qwx  & \qw \qwx  & \qw \qwx  & \qw \qwx  & \qw \qwx  & \onecontrol\qwx  & \qw & \qw & \qw & \qw & \qw & \qw \\
\lstick{\ket{q_{11}}} & \qw & \qw & \qw & \qw & \qw & \qw  & \gate{X_{+1}}\qwx  & \twocontrol\qwx  & \qw \qwx  & \twocontrol\qwx  & \gate{X_{-1}}\qwx  & \qw  & \qw & \qw & \qw & \qw & \qw & \qw \\
\lstick{\ket{q_{12}}} & \qw & \qw & \qw & \qw & \qw & \onecontrol & \qw \qwx  & \qw  & \qw \qwx  & \qw  & \qw \qwx  & \onecontrol & \qw & \qw & \qw & \qw & \qw & \qw \\
\lstick{\ket{q_{13}}} & \qw & \qw & \qw & \qw & \qw & \gate{X_{+1}}\qwx  & \twocontrol\qwx  & \qw  & \qw \qwx  & \qw  & \twocontrol\qwx  & \gate{X_{-1}}\qwx  & \qw & \qw & \qw & \qw & \qw & \qw \\
\lstick{\ket{q_{14}}} & \qw & \qw & \qw & \qw & \qw & \onecontrol\qwx  & \qw  & \qw  & \qw \qwx  & \qw  & \qw  & \onecontrol\qwx  & \qw & \qw & \qw & \qw & \qw & \qw \\
\lstick{\ket{q_{15}}} & \qw & \qw & \qw & \qw & \qw & \qw  & \qw  & \qw  & \gate{U}\qwx  & \qw  & \qw  & \qw  & \qw & \qw & \qw & \qw & \qw & \qw \\
}
\]

%% file: txt/5applications_to_algorithms.tex
\section{Application to Algorithms} \label{sec:application to algorithms}
The Generalized Toffoli gate is an important primitive in a broad range of quantum algorithms. In this section, we survey some of the applications of our circuit decomposition.

\subsection{Artificial Quantum Neuron}
The artificial quantum neuron \cite{ArtificialQuantumNeuron} is a promising target application for our circuit construction, because the algorithm's circuit implementation is dominated by large Generalized Toffoli gates. The algorithm may exhibit an exponential advantage over classical perceptron encoding and it has already been executed on current quantum hardware. Moreover, the threshold behavior of perceptrons has inherent noise resilience, which makes the artificial quantum neuron particularly promising as a near-term application on noisy systems. The current implementation of the neuron on IBM quantum computers relies on ancilla qubits \cite{ArtificialQuantumNeuron_PrivateComm} which constrains the circuit width to $N=4$ data qubits. Our circuit construction offers a path to larger circuit sizes without waiting for larger hardware.


\subsection{Grover's Algorithm}
\begin{figure}[ht]
  \input{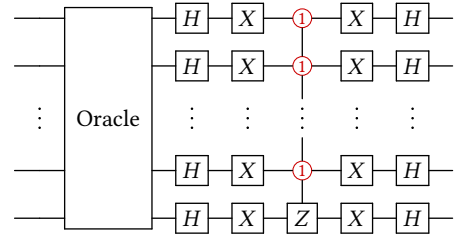}
  \caption{Each iteration of Grover Search has a multiply-controlled $Z$ gate. Our logarithmic depth decomposition, reduces a $\log M$ factor in Grover's algorithm to $\log\log M$.}
  \label{fig:grover_search}
\end{figure}

Grover's Algorithm for search over $M$ unordered items requires just $O(\sqrt{M})$ oracle queries. However, each oracle query is followed by a post-processing step which requires a multiply-controlled gate with $N = \lceil \log_2 M \rceil$ controls \cite{Nielsen}. The explicit circuit diagram is shown in Figure~\ref{fig:grover_search}.

Our log-depth circuit construction directly applies to the multiply-controlled gate. Thus, we reduce a $\log M$ factor in Grover search's time complexity to  $\log \log M $ via our ancilla-free qutrit decomposition.

\subsection{Incrementer} \label{sec:incrementer}
The Incrementer circuit performs the $+ 1 \mod 2^N$ operation to a register of $N$ qubits. While logarithmic circuit depth can be achieved with linear ancilla qubits \cite{Draper}, the best ancilla-free incrementers require either linear depth with large linearity constants \cite{Gidney} or quadratic depth \cite{Barenco}. Using alternate control activations for our Generalized Toffoli gate decomposition, the incrementer circuit is reduced to $O(\log^2 N)$ depth with no ancilla, a significant improvement over past work.

\begin{figure}[ht]
  \input{figs/BTBPlusOne_with_outputs.qcircuit}
  \caption{Our circuit decomposition for the Incrementer. At each subcircuit in the recursive design, multiply-controlled gates are used to efficiently propagate carries over half of the subcircuit. The $\ket{2}$ control checks for carry generation and the chain of $\ket{1}$ controls checks for carry propagation. The circuit depth is $\log^2 N$, which is only possible because of our log depth multiply-controlled gate primitive.}
  \label{fig:incrementer}
\end{figure}
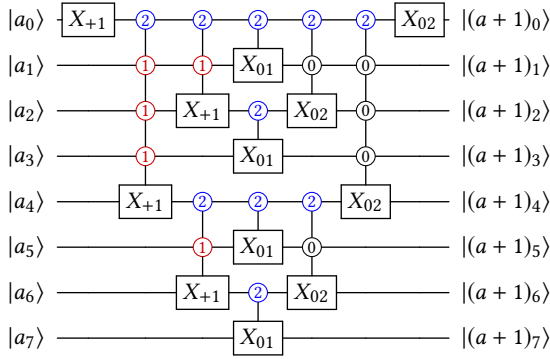

Our incrementer circuit construction is shown in Figure~\ref{fig:incrementer} for an $N=8$ wide register. The multiple-controlled $X_{+1}$ gates perform the job of computing carries: a carry is performed iff the least significant bit generates (represented by the $\ket{2}$ control) and all subsequent bits propagate (represented by the consecutive $\ket{1}$ controls). We present an $N=8$ incrementer here and have verified the general construction, both by formal proof and by explicit circuit simulation for larger $N$.

The critical path of this circuit is the chain of $\log{N}$ multiply-controlled gates (of width $\frac{N}{2}$, $\frac{N}{4}$, $\frac{N}{8}$, ...) which act on $\ket{a_0}$. Since our  multiply-controlled gate decomposition has log-depth, we arrive at a total circuit depth circuit scaling of $\log^2 N$.

\subsection{Arithmetic Circuits and Shor's Algorithm}
The Incrementer circuit is a key subcircuit in many other arithmetic circuits such as constant addition, modular multiplication, and modular exponentiation. Further, the modular exponentiation circuit is the bottleneck in the runtime for executing Shor's algorithm for factorization \cite{Gidney, Microsoft}. While a shallower Incrementer circuit alone is not sufficient to reduce the asymptotic cost of modular exponentiation (and therefore Shor's algorithm), it does reduce constants relative to qubit-only circuits.

\subsection{Error Correction and Fault Tolerance}
The Generalized Toffoli gate has applications to circuits for both error correction \cite{Cory} and fault tolerance \cite{Dennis}. We foresee two paths of applying these circuits. First, our circuit construction can be used to construct error-resilient \textit{logical qubits} more efficiently. This is critical for quantum algorithms like Grover's and Shor's which are expected to require such logical qubits. In the nearer-term, NISQ algorithms are likely to make use of limited error correction. For instance, recent results have demonstrated that error correcting a single qubit at a time for the Variational Quantum Eigensolver algorithm can significantly reduce total error \cite{Otten}. Thus, our circuit construction is also relevant for NISQ-era error correction.

%% file: figs/BTBPlusOne_with_outputs.qcircuit

%

\[
\Qcircuit @R=0.5em @C=0.2em {
\lstick{\ket{a_{0}}} & \gate{X_{+1}} & \twocontrol & \twocontrol & \twocontrol & \twocontrol & \twocontrol & \gate{X_{02}} & \rstick{\ket{(a+1)_{0}}} \qw & \\
\lstick{\ket{a_{1}}} & \qw  & \onecontrol\qwx  & \onecontrol\qwx  & \gate{X_{01}}\qwx  & \qcontrol{0}\qwx  & \qcontrol{0}\qwx  & \qw  & \rstick{\ket{(a+1)_{1}}} \qw & \\
\lstick{\ket{a_{2}}} & \qw  & \onecontrol\qwx  & \gate{X_{+1}}\qwx  & \twocontrol & \gate{X_{02}}\qwx  & \qcontrol{0}\qwx  & \qw  & \rstick{\ket{(a+1)_{2}}} \qw & \\
\lstick{\ket{a_{3}}} & \qw  & \onecontrol\qwx  & \qw  & \gate{X_{01}}\qwx  & \qw  & \qcontrol{0}\qwx  & \qw  & \rstick{\ket{(a+1)_{3}}} \qw & \\
\lstick{\ket{a_{4}}} & \qw  & \gate{X_{+1}}\qwx  & \twocontrol & \twocontrol & \twocontrol & \gate{X_{02}}\qwx  & \qw  & \rstick{\ket{(a+1)_{4}}} \qw & \\
\lstick{\ket{a_{5}}} & \qw  & \qw  & \onecontrol\qwx  & \gate{X_{01}}\qwx  & \qcontrol{0}\qwx  & \qw  & \qw  & \rstick{\ket{(a+1)_{5}}} \qw & \\
\lstick{\ket{a_{6}}} & \qw  & \qw  & \gate{X_{+1}}\qwx  & \twocontrol & \gate{X_{02}}\qwx  & \qw  & \qw  & \rstick{\ket{(a+1)_{6}}} \qw & \\
\lstick{\ket{a_{7}}} & \qw  & \qw  & \qw  & \gate{X_{01}}\qwx  & \qw  & \qw  & \qw  & \rstick{\ket{(a+1)_{7}}} \qw & \\
}
\]

%% file: txt/6simulator.tex
\section{Simulator} \label{sec:simulator}
To simulate our circuit constructions, we developed a qudit simulation library, built on Google's Cirq Python library \cite{Cirq}. Cirq is a qubit-based quantum circuit library and includes a number of useful abstractions for quantum states, gates, circuits, and scheduling.

Our work extends Cirq by discarding the assumption of two-level qubit states. Instead, all state vectors and gate matrices are expanded to apply to $d$-level qudits, where $d$ is a circuit parameter. We include a library of common gates for $d=3$ qutrits. Our software adds a comprehensive noise simulator, detailed below in Section~\ref{subsec:noise simulation}.

In order to verify our circuits are logically correct, we first simulated them with noise disabled. We extended Cirq to allow gates to specify their action on classical non-superposition input states without considering full state vectors. Therefore, each classical input state can be verified in space and time proportional to the circuit width. By contrast, Cirq's default simulation procedure relies on a dense state vector representation requiring space and time exponential in the circuit width. Reducing this scaling from exponential to linear dramatically improved our verification procedure, allowing us to verify circuit constructions for all possible classical inputs across circuit sizes up to widths of 14.

Our software is fully open source \cite{QutritsGithub}.

\subsection{Noise Simulation} \label{subsec:noise simulation}
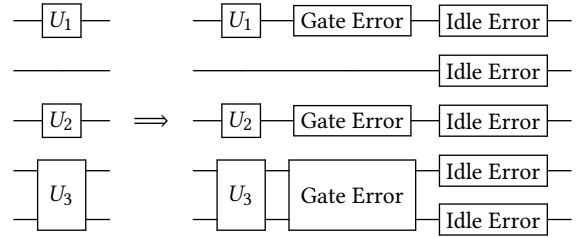
\begin{figure}[ht]
  \input{figs/noise_simulation_model.qcircuit}
  \caption{This \texttt{Moment} comprises three gates executed in parallel. To simulate with noise, we first apply the ideal gates, followed by a gate error noise channel on each affected qudit. This gate error noise channel depends on whether the corresponding gate was single- or two- qudit. Finally, we apply an idle error to every qudit. The idle error noise channel depends on the duration of the \texttt{Moment}.}
  \label{fig:noise_simulation_methodology}
\end{figure}

Figure~\ref{fig:noise_simulation_methodology} depicts a schematic view of our noise simulation procedure which accounts for both gate errors and idle errors, described below. To determine when to apply each gate and idle error, we use Cirq's scheduler which schedules each gate as early as possible, creating a sequence of \texttt{Moment}'s of simultaneous gates. During each \texttt{Moment}, our noise simulator applies a gate error to every qudit acted on. Finally, the simulator applies an idle error to every qudit. This noise simulation methodology is consistent with previous simulation techniques which have accounted for either gate errors \cite{Miller} or idle errors \cite{QX_Simulator}.

Gate errors arise from the imperfect application of quantum gates. Two-qudit gates are noisier than single-qudit gates \cite{IBMDevices}, so we apply different noise channels for the two. Our specific gate error probabilities are given in Section~\ref{sec:error modeling}.

Idle errors arise from the continuous decoherence of a quantum system due to energy relaxation and interaction with the environment. The idle errors differ from gate errors in two ways which require special treatment:
\begin{enumerate}
    \item Idle errors depend on duration, which in turn depend on the schedule of simultaneous gates (\texttt{Moment}s). In particular, two-qudit gates take longer to apply than single-qudit gates. Thus, if a \texttt{Moment} contains a two-qudit gate, the idling errors must be scaled appropriately. Our specific scaling factors are given in Section~\ref{sec:error modeling}.
    \item For the generic model of gate errors, the error channel is applied with probability independent of the quantum state. This is not true for idle errors such as $T_1$ amplitude damping, which only applies when the qudit is in an excited state. This is treated in the simulator by computing idle error probabilities during each \texttt{Moment}, for each qutrit.
\end{enumerate}
Gate errors are reduced by performing fewer \textit{total gates}, and idle errors are reduced by decreasing the circuit \textit{depth}. Since our circuit constructions asymptotically decrease the depth, this means our circuit constructions scale favorably in terms of asymptotically fewer idle errors.

Our full noise simulation procedure is summarized in Algorithm ~\ref{alg:pseudocode}. The ultimate metric of interest is the mean \textit{fidelity}, which is defined as the squared overlap between the ideal (noise-free) and actual output state vectors. Fidelity expresses the probability of overall successful execution. We do not consider initialization errors and readout errors, because our circuit constructions maintain binary input and output, only occupying the qutrit $\ket{2}$ states during intermediate computation. Therefore, the initialization and readout errors for our circuits are identical to those for conventional qubit circuits.

\SetEndCharOfAlgoLine{}
\begin{algorithm}
\SetAlgoLined
$\ket{\Psi} \leftarrow$ random initial state vector\;
$\ket{\Psi}_{\text{ideal}} =$ circuit applied to $\ket{\Psi}$ without noise\;
\BlankLine
\ForEach{\textup{\texttt{Moment}}}{
 \ForEach{\textup{\texttt{Gate} $\in$ \texttt{Moment}}} {
   $\ket{\psi} \leftarrow$ \texttt{Gate} applied to $\ket{\psi}$\;
   \texttt{GateError} $\leftarrow$ DrawRand(\texttt{GateError Prob.})\;
   $\ket{\psi} \leftarrow$ \texttt{GateError} applied to $\ket{\psi}$\;
 }
 \BlankLine
 \ForEach{\textup{\texttt{Qutrit}}} {
   \eIf{\textup{\texttt{Moment}} has 2-qudit gate}{
     \texttt{IdleErrors} $\leftarrow$ long-duration idle errors\;
   }{
     \texttt{IdleErrors} $\leftarrow$ short-duration idle errors\;
  }
  \texttt{Prob.} $\leftarrow [ \|\texttt{M} \ket{\Psi}\|^2 \text{ for } \texttt{M} \in \texttt{IdleErrors}]$ \;
  \texttt{IdleError} $\leftarrow$ DrawRand(\texttt{Prob.})\;
  $\ket{\psi} \leftarrow$ \texttt{IdleError} applied to $\ket{\psi}$\;
  Renormalize($\ket{\psi}$)\;
 }
}
\Return $\braket{\Psi_{\text{ideal}} | \Psi}^2,$ fidelity between ideal \& actual output;
 \caption{Pseudocode for each simulation trial, given a particular circuit and noise model.}
   \label{alg:pseudocode}
\end{algorithm}

We also do not consider crosstalk errors, which occur when gates are executed in parallel. The effect of crosstalk is very device-dependent and difficult to generalize. Moreover, crosstalk can be mitigated by breaking each \texttt{Moment} into a small number of sub-moments and then scheduling two-qutrit operations to reduce crosstalk, as demonstrated in prior work \cite{Crosstalk1, Crosstalk2}.

\subsection{Simulator Efficiency}
Simulating a quantum circuit with a classical computer is, in general, exponentially difficult in the size of the input because the state of $N$ qudits is represented by a state vector of $d^N$ complex numbers. For 14 qutrits, with complex numbers stored as two 8-byte floats (\texttt{complex128} in NumPy), a state vector occupies 77 megabytes.

A naive circuit simulation implementation would treat every quantum gate or \texttt{Moment} as a $d^N \times d^N$ matrix. For 14 qutrits, a single such matrix would occupy 366 terabytes--out of range of simulability. While the exponential nature of simulating our circuits is unavoidable, we mitigate the cost by using a variety of techniques which rely only on state vectors, rather than full square matrices. For example, we maintain Cirq's approach of applying gates by Einstein Summation \cite{TensorNetworks}, which obviates computation of the $d^N \times d^N$ matrix corresponding to every gate or \texttt{Moment}.

Our noise simulator only relies on state vectors, by adopting the quantum trajectory methodology \cite{QuantumTrajectories, CppQuantumTrajectories}, which is also used by the Rigetti PyQuil noise simulator \cite{RigettiPyQuil}. At a high level, the effect of noise channels like gate and idle errors is to turn a coherent quantum state into an incoherent mix of classical probability-weighted quantum states (for example, $\ket{0}$ and $\ket{1}$ with 50\% probability each). The most complete description of such an incoherent quantum state is called the density matrix and has dimension $d^N \times d^N$. The quantum trajectory methodology is a stochastic approach--instead of maintaining a density matrix, only a single state is propagated and the error term is drawn randomly at each timestep. Over repeated trials, the quantum trajectory methodology converges to the same results as from full density matrix simulation \cite{RigettiPyQuil}. Our simulator employs this technique--each simulation in Algorithm~\ref{alg:pseudocode} constitutes a single quantum trajectory trial. At every step, a specific \texttt{GateError} or \texttt{IdleError} term is picked, based on a weighted random draw.

Finally, our random state vector generation function was also implemented in $O(d^N)$ space and time. This is an improvement over other open source libraries \cite{QuTiP, QuTiP2}, which perform random state vector generation by generating full $d^N \times d^N$ unitary matrices from a Haar-random distribution and then truncating to a single column. Our simulator directly computes the first column and circumvents the full matrix computation.

With optimizations, our simulator is able to simulate circuits up to 14 qutrits in width. This is in the range as other state-of-the-art noisy quantum circuit simulations \cite{28QubitNoisySimulation} (since 14 qutrits $\approx$ 22 qubits). While each simulation trial took several minutes (depending on the particular circuit and noise model), we were able to run trials in parallel over multiple processes and multiple machines, as described in Section~\ref{sec:results}.

%% file: figs/noise_simulation_model.qcircuit

%

\[
\Qcircuit @R=0.75em @C=1em {
\lstick{} & \gate{U_1} & \qw &    &    & \gate{U_1} & \gate{\mbox{Gate Error}} & \gate{\mbox{Idle Error}} & \qw \\
\lstick{} & \qw  & \qw &    &    & \qw  & \qw  & \gate{\mbox{Idle Error}} & \qw \\
\lstick{} & \gate{U_2} & \qw & \push{\Longrightarrow} &    & \gate{U_2} & \gate{\mbox{Gate Error}} & \gate{\mbox{Idle Error}} & \qw \\
\lstick{} & \multigate{1}{U_3} & \qw &    &    & \multigate{1}{U_3} & \multigate{1}{\mbox{Gate Error}} & \gate{\mbox{Idle Error}} & \qw \\
\lstick{} & \ghost{U_3} & \qw &    &    & \ghost{U_3} & \ghost{\mbox{Gate Error}} & \gate{\mbox{Idle Error}} & \qw \\
}
\]

%% file: txt/7noise_models.tex
\section{Noise Models} \label{sec:error modeling}
In this section, we describe our noise models at a high level, with mathematical details described in Appendix~\ref{app:detailed_noise_models}. We chose noise models which represent realistic near-term machines. We first present a generic, parametrized noise model roughly applicable to all quantum systems. We then present specific parameters, under the generic noise model, which apply to near-term superconducting quantum computers. Finally, we present a specific noise model for trapped ion quantum computers.

\subsection{Generic Noise Model} \label{subsec:generic}
\subsubsection{Gate Errors}
The scaling of gate errors for a $d$-level qudit can be roughly summarized as increasing as $d^4$ for two-qudit gates and $d^2$ for single-qudit gates. For $d=2$, there are 4 single-qubit gate error channels and 16 two-qubit gate error channels. For $d=3$ there are 9 and 81 single- and two- qutrit gate error channels respectively. Consistent with other simulators \cite{RigettiPyQuil, QX_Simulator}, we use the symmetric depolarizing gate error model, which assumes equal probabilities between each error channel. Under these noise models, two-qutrit gates are $(1-80p_2) / (1-15p_2)$ times less reliable than two-qubit gates, where $p_2$ is the probability of each two-qubit gate error channel. Similarly, single-qutrit gates are $(1-8p_1) / (1 - 3p_1)$ times less reliable than single-qubit gates, where $p_1$ is the probability of each single-qubit gate error channel.

\subsubsection{Idle Errors}
\label{subsubsec:idle errors}
Our treatment of idle errors focuses on the relaxation from higher to lower energy states in quantum devices. This is called amplitude damping or $T_1$ relaxation. This noise channel irreversibly takes qudits to lower states. For qubits, the only amplitude damping channel is from $\ket{1}$ to $\ket{0}$, and we denote this damping probability as $\lambda_1$. For qutrits, we also model damping from $\ket{2}$ to $\ket{0}$, which occurs with probability $\lambda_2$.

\subsection{Superconducting QC}
\label{subsec:superconducting}
We chose four noise models based on superconducting quantum computers expected in the next few years. These noise models comply with the generic noise model above and are thus parametrized by $p_1$, $p_2$, $\lambda_1$ and $\lambda_2$. The $\lambda_i$ probabilities are derived from two other experimental parameters: the gate time $\Delta t$ and $T_1$, a timescale that captures how long a qudit persists coherently.

As a starting point for representative near-term noise models, we consider parameters for \textit{current} superconducting quantum computers. For IBM's public cloud-accessible superconducting quantum computers, we have $3p_1 \approx 10^{-3}$ and $15p_2 \approx 10^{-2}$. The duration of single- and two- qubit gates is $\Delta t \approx 100ns$ and $\Delta t \approx 300ns$ respectively, and the IBM devices have $T_1 \approx 100 \mu s$ \cite{IBMDevices, Linke}.

However, simulation for these current parameters indicates an error is almost certain to occur during execution of a modest size 14-input Generalized Toffoli circuit. This motivates us to instead consider noise models for better devices which are a few years away. Accordingly, we adopt a baseline superconducting noise model, labeled as \textcolor{brown}{SC}, corresponding to a superconducting device which has 10x lower gate errors and 10x longer $T_1$ duration than the current IBM hardware. This range of parameters has already been achieved experimentally in superconducting devices for gate errors \cite{LowGateError1, LowGateError2} and for $T_1$ duration \cite{LongT1_1, LongT1_2} independently. Faster gates (shorter $\Delta t$) are yet another path towards greater noise resilience. We do not vary gate speeds, because errors only depend on the $\Delta t / T_1$ ratio, and we already vary $T_1$. In practice however, faster gates could also improve noise-resilience.

We also consider three additional near-term device noise models, indexed to the \textcolor{brown}{SC} noise model. These three models further improve gate errors, $T_1$, or both, by a 10x factor. The specific parameters are given in Table~\ref{tab:sc_noise_models}. Our 10x improvement projections are realistic extrapolations of progress in hardware. In particular, Schoelkopf's Law--the quantum analogue of Moore's Law--has observed that $T_1$ durations have increased by 10x every 3 years for the past 20 years \cite{LesHouchesQEDNotes}. Hence, 100x longer $T_1$ is a reasonable projection for devices that are $\sim 6$ years away.

\begin{table}[]
\centering
\begin{tabular}{l|ll|l}
  Noise Model &  $3 p_1$ & $15 p_2$ & $T_1$ \\ \hline
 \textcolor{brown}{SC} & $10^{-4}$ & $10^{-3}$ & 1 ms \rule{0pt}{2.6ex}
 \\ \textcolor{brown}{SC+T1} & $10^{-4}$ & $10^{-3}$ &  10 ms
 \\ \textcolor{brown}{SC+GATES} & $10^{-5}$ & $10^{-4}$ & 1 ms
 \\ \textcolor{brown}{SC+T1+GATES} & $10^{-5}$ & $10^{-4}$ & 10 ms
\end{tabular}
\caption{Noise models simulated for superconducting devices. Current publicly accessible IBM superconducting quantum computers have single- and two- qubit gate errors of $3p_1 \approx 10^{-3}$ and $15p_2 \approx 10^{-2}$, as well as $T_1$ lifetimes of 0.1 ms \cite{IBMDevices, Linke}. Our baseline benchmark, \textcolor{brown}{SC}, assumes 10x better gate errors and $T_1$. The other three benchmarks add a further 10x improvement to $T_1$, gate errors, or both.}
\label{tab:sc_noise_models}
\end{table}

\subsection{Trapped Ion $^{171}$Yb$^+$ QC}
\label{subsec:trapped ion}
We also simulated noise models for trapped ion quantum computing devices.  Trapped ion devices are well matched to our qutrit-based circuit constructions because they feature all-to-all connectivity \cite{brown2016co}, and many ions that are ideal candidates for QC devices are naturally multi-level systems.

We focus on the $^{171}$Yb${^+}$ ion, which has been experimentally demonstrated as both a qubit and qutrit \cite{HesingA, HesingB}. Trapped ions are often favored in QC schemes due to their long $T_1$ times. One of the main advantages of using a trapped ion is the ability to take advantage of magnetically insensitive states known as "clock states." By defining the computational subspace on these clock states, idle errors caused from fluctuations in the magnetic field are minimized--this is termed a \textcolor{brown}{DRESSED\_QUTRIT}, in contrast with a \textcolor{brown}{BARE\_QUTRIT}. However, compared to superconducting devices, gates are much slower. Thus, gate errors are the dominant error source for ion trap devices. We modelled a fundamental source of these errors: the spontaneous scattering of photons originating from the lasers used to drive the gates. The duration of single- and two- qubit gates used in this calculation was $\Delta t \approx 1$ $\mu$s and $\Delta t \approx 200$ $\mu$s respectively \cite{PhysRevA.97.052301}. The single- and two- qudit gate error probabilities are given in Table~\ref{tab:ti_noise_models}.

\begin{table}[]
\centering
\begin{tabular}{l|ll}
  Noise Model &  $p_1$ & $p_2$ \\ \hline
 \textcolor{brown}{TI\_QUBIT} & $6.4 \times 10^{-4}$ & $1.3 \times 10^{-4}$  \rule{0pt}{2.6ex}
 \\ \textcolor{brown}{BARE\_QUTRIT} & $2.2 \times 10^{-4}$ & $4.3 \times 10^{-4}$
 \\ \textcolor{brown}{DRESSED\_QUTRIT} & $1.5 \times 10^{-4}$ & $3.1 \times 10^{-4}$
\end{tabular}
\caption{Noise models simulated for trapped ion devices. The single- and two- qutrit gate error channel probabilities are based on calculations from experimental parameters. For all three models, we use single- and two- qudit gate times of $\Delta t \approx 1$ $\mu s$ and $\Delta t \approx 200$ $\mu s$ respectively.}
\label{tab:ti_noise_models}
\end{table}

%% file: txt/8results.tex
\section{Results} \label{sec:results}
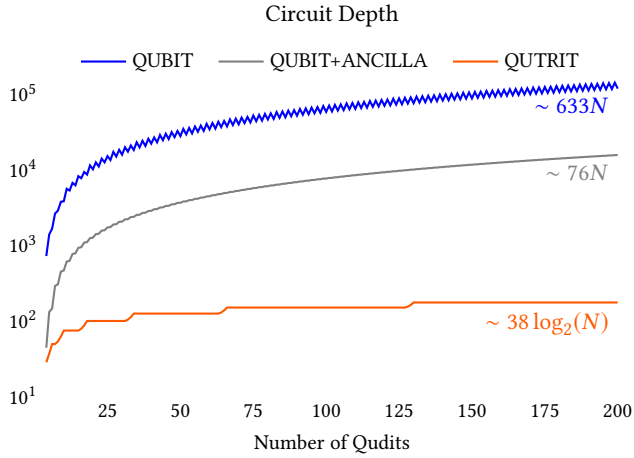
\begin{figure}[ht]
    \input{figs/graph_circuit_depth.tikz}
    \caption{Exact circuit depths for all three benchmarked circuit constructions for the N-controlled Generalized Toffoli up to $N=200$. Both QUBIT and QUBIT+ANCILLA scale linearly in depth and both are bested by QUTRIT's logarithmic depth. }
    \label{fig:circuit_depths}
\end{figure}

Figure~\ref{fig:circuit_depths} plots the exact circuit depths for all three benchmarked circuits. The qubit-based circuit constructions from past work are linear in depth and have a high linearity constant. Augmenting with a single borrowed ancilla reduces the circuit depth by a factor of 8. However, both circuit constructions are surpassed significantly by our qutrit construction, which scales logarithmically in $N$ and has a relatively small leading coefficient.

\begin{figure}[ht]
    \input{figs/graph_2gate_count.tikz}
    \caption{Exact two-qudit gate counts for the three benchmarked circuit constructions for the N-controlled Generalized Toffoli. All three plots scale linearly; however the QUTRIT construction has a substantially lower linearity constant.}
    \label{fig:two_qudit_gate_counts}
\end{figure}
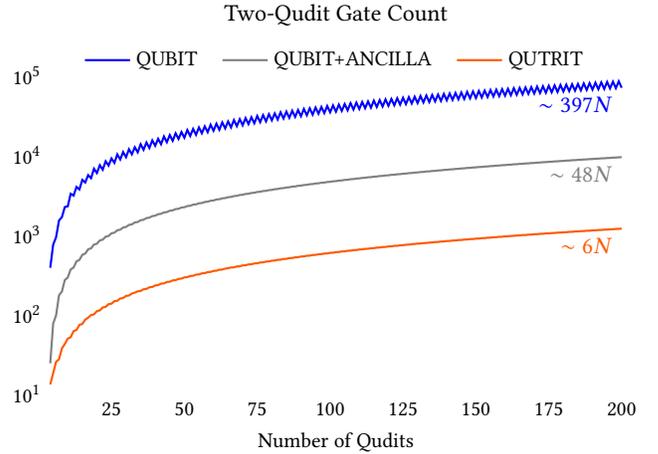

Figure~\ref{fig:two_qudit_gate_counts} plots the total number of two-qudit gates for all three circuit constructions. As noted in Section~\ref{sec:circuit constructions}, our circuit construction is not asymptotically better in total gate count--all three plots have linear scaling. However, as emphasized by the logarithmic vertical axis, the linearity constant for our qutrit circuit is 70x smaller than for the equivalent ancilla-free qubit circuit and 8x smaller than for the borrowed-ancilla qubit circuit.

Our simulations under realistic noise models were run in parallel on over 100 n1-standard-4 Google Cloud instances. These simulations represent over 20,000 CPU hours, which was sufficient to estimate mean fidelity to an error of $2 \sigma < 0.1\%$ for each circuit-noise model pair.

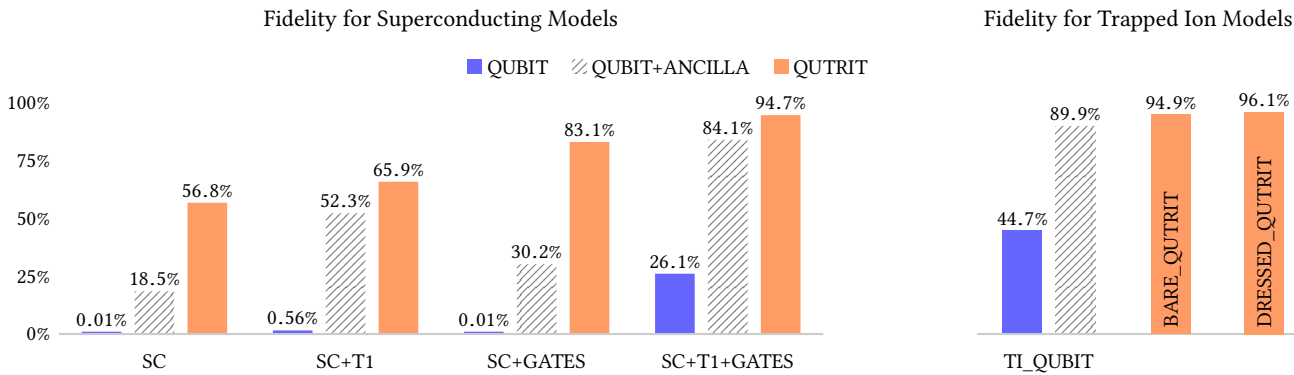
\begin{figure*}[ht]
    \centering
        \input{figs/graph_simulation_sc.tikz}%
        \:\:\:\:\:\:\:\:\:\:\:\:\:\:\:%
        \input{figs/graph_simulation_ti.tikz}
    \caption{Circuit simulation results for all possible pairs of circuit constructions and noise models. Each bar represents 1000+ trials, so the error bars are all $2\sigma < 0.1\%$. Our QUTRIT construction significantly outperforms the QUBIT construction. The QUBIT+ANCILLA bars are drawn with dashed lines to emphasize that it has access to an extra ancilla bit, unlike our construction.}
    \label{fig:simulation_results}
\end{figure*}

The full results of our circuit simulations are shown in Figure~\ref{fig:simulation_results}. All simulations are for the 14-input (13 controls, 1 target) Generalized Toffoli gate. We simulated each of the three circuit benchmarks against each of our noise models (when applicable), yielding the 16 bars in the figure.

%% file: figs/graph_circuit_depth.tikz
\begin{tikzpicture}
\pgfplotsset{every axis/.append style={thick}}
\pgfplotsset{every tick label/.append style={font=\small}}
\pgfplotsset{every axis label/.append style={font=\small}}

\begin{semilogyaxis}[
    title=Circuit Depth,
    xlabel=Number of Qudits,
    width=\columnwidth + 20pt,
    height=175pt,
    xmin=4, xmax=200,
    ymin=10,
    restrict x to domain=4:200,
    xtick distance=25,
    ,
    legend style={draw=none, fill=none, at={(0.5,1.03)},anchor=north,font=\small},
    legend columns=-1,
    axis line style={draw=none},
    tick style={draw=none},
    clip=false,
]

\addplot[color=\colorQubit] table[x=Qubits, y=Depth, col sep=comma]
    {data/gateCounts-CnXLinear.csv}
node [anchor=north east] {$\sim 633N$};
\addlegendentry{QUBIT\:\:\:\:};

\addplot[color=\colorQubitBB] table[x=Qubits, y=Depth, col sep=comma]
    {data/gateCounts-CnXLinearBB.csv}
node [anchor=north east] {$\sim 76N$};
\addlegendentry{QUBIT+ANCILLA\:\:\:\:};

\addplot[color=\colorQutrit] table[x=Qubits, y=Depth, col sep=comma]
    {data/gateCounts-CnXQutrit.csv}
node [anchor=north east] {$\sim 38\log_2(N)$};
\addlegendentry{QUTRIT};

\end{semilogyaxis}
\end{tikzpicture}

%% file: figs/graph_2gate_count.tikz
\begin{tikzpicture}
\pgfplotsset{every axis/.append style={thick}}
\pgfplotsset{every tick label/.append style={font=\small}}
\pgfplotsset{every axis label/.append style={font=\small}}

\begin{semilogyaxis}[
    title=Two-Qudit Gate Count,
    xlabel=Number of Qudits,
    width=\columnwidth + 20pt,
    height=175pt,
    xmin=4, xmax=200,
    ymin=10,
    restrict x to domain=4:200,
    xtick distance=25,
    ,
    legend style={draw=none, fill=none, at={(0.5,1.03)},anchor=north,font=\small},
    legend columns=-1,
    axis line style={draw=none},
    tick style={draw=none},
    clip=false,
]

\addplot[color=\colorQubit] table[x=Qubits, y=Two Qubit Gate Count, col sep=comma]
    {data/gateCounts-CnXLinear.csv}
node [anchor=north east] {$\sim 397N$};
\addlegendentry{QUBIT\:\:\:\:};

\addplot[color=\colorQubitBB] table[x=Qubits, y=Two Qubit Gate Count, col sep=comma]
    {data/gateCounts-CnXLinearBB.csv}
node [anchor=north east] {$\sim 48N$};
\addlegendentry{QUBIT+ANCILLA\:\:\:\:};

\addplot[color=\colorQutrit] table[x=Qubits, y=Two Qubit Gate Count, col sep=comma]
    {data/gateCounts-CnXQutrit.csv}
node [anchor=north east] {$\sim 6N$};
\addlegendentry{QUTRIT};

\end{semilogyaxis}
\end{tikzpicture}

%% file: figs/graph_simulation_sc.tikz
\begin{tikzpicture}[baseline]
\pgfplotsset{every axis/.append style={thick}}
\pgfplotsset{every tick label/.append style={font=\small}}
\pgfplotsset{every axis label/.append style={font=\small}}

\newcommand{\extraBarHeight}{1}

\begin{axis}[
    title=Fidelity for Superconducting Models,
    symbolic x coords={left,SC,a,SC+T1,b,SC+GATES,c,SC+T1+GATES,right},
    width=.66\textwidth,
    height=150pt,
    ybar=5pt,
    bar width=15pt,
    xmin=left, xmax=right,
    ymin=0, ymax=120,
    ytick={0, 25, 50, 75, 100},
    ytick distance=25,
    xtick=data,
    ,
    legend style={draw=none, fill=none, at={(0.8,1.03)},anchor=north,font=\small},
    legend columns=-1,
    legend image code/.code={\draw[#1, draw=none] (0em,-0.2em) rectangle (0.6em,0.4em);},
    axis line style={draw=black!20!white},
    axis on top,
    y axis line style={draw=none},
    axis x line*=bottom,
    tick style={draw=none},
    yticklabel={\pgfmathparse{\tick*1}\pgfmathprintnumber{\pgfmathresult}\%},
    clip=false,
    enlarge y limits=0,
    ,
    nodes near coords always on top/.style={
        every node near coord/.append style={
            anchor=south,
            rotate=0,
            font=\small,
            inner sep=0.2em,
        },
    },
    nodes near coords={
        \StrPosition{\pgfplotspointmeta}{Y}[\Result]%
        \StrGobbleLeft{\pgfplotspointmeta}{\Result}[\Result]%
        \StrGobbleRight{\Result}{1}[\Result]%
        \pgfmathparse{\Result<10}%
        \ifnum\pgfmathresult=1
            \pgfmathparse{\Result-\extraBarHeight}%
        \else
            \pgfmathparse{\Result}%
        \fi%
        \pgfmathprintnumber[fixed,precision=2]{\pgfmathresult}\%
    },
    nodes near coords always on top,
]

\addplot[style={color=transparent, draw=none, fill=\colorQubitLight, mark=none}] coordinates {
    (SC, 1.01)  
    (SC+T1, 1.56)  
    (SC+GATES, 1.01)  
    (SC+T1+GATES, 26.1)  
};
\addlegendentry{QUBIT\:\:\:\:};

\addplot[style={draw=none, pattern color=\colorQubitBB, pattern=north east lines, mark=none}] coordinates {
    (SC, 18.5)  
    (SC+T1, 52.3)  
    (SC+GATES, 30.2)  
    (SC+T1+GATES, 84.1)  
};
\addlegendentry{QUBIT+ANCILLA\:\:\:\:};

\addplot[style={draw=none, fill=\colorQutritLight, mark=none}] coordinates {
    (SC, 56.8)  
    (SC+T1, 65.9)  
    (SC+GATES, 83.1)  
    (SC+T1+GATES, 94.7)  
};
\addlegendentry{QUTRIT};

\end{axis}

%
%
%
%
\end{tikzpicture}%

%% file: figs/graph_simulation_ti.tikz
\begin{tikzpicture}[baseline]
\pgfplotsset{every axis/.append style={thick}}
\pgfplotsset{every y tick label/.append style={font=\small}}
\pgfplotsset{every x tick label/.append style={font=\small}}
\pgfplotsset{every axis label/.append style={font=\small}}

\newcommand{\beginCustomBarAxisFirst}[1]{
\begin{axis}[
    title=Fidelity for Trapped Ion Models,
    symbolic x coords={#1},
    width=0.33\textwidth,
    height=150pt,
    ybar=5pt,
    bar width=15pt,
    xmin=left, xmax=right,
    ymin=0, ymax=120,
    ytick={0, 25, 50, 75, 100},
    ytick distance=25,
    xtick=data,
    enlarge x limits=0,
    ,
    legend style={draw=none, fill=none, at={(0.5,1.03)},anchor=north,font=\small},
    legend columns=2,
    legend image code/.code={\draw[##1, draw=none] (0em,-0.2em) rectangle (0.6em,0.4em);},
    axis line style={draw=black!20!white},
    axis on top,
    y axis line style={draw=none},
    axis x line*=bottom,
    tick style={draw=none},
    yticklabel={~},
    clip=false,
    enlarge y limits=0,
    ,
    nodes near coords always on top/.style={
        every node near coord/.append style={
            anchor=south,
            rotate=0,
            font=\small,
            inner sep=0.2em,
        },
    },
    nodes near coords={
        \pgfmathprintnumber[fixed,fixed zerofill,precision=1]{\pgfplotspointmeta}\%
    },
    nodes near coords always on top,
]
}

\newcommand{\beginCustomBarAxis}[1]{
\begin{axis}[
    title=,
    symbolic x coords={#1},
    width=0.33\textwidth,
    height=150pt,
    ybar=5pt,
    bar width=15pt,
    xmin=left, xmax=right,
    ymin=0, ymax=120,
    ytick={0, 25, 50, 75, 100},
    ytick distance=25,
    xtick=data,
    enlarge x limits=0,
    x tick label style={rotate=90, anchor=west, yshift=-0.1em, xshift=0.4em, color=black},
    ,
    legend style={draw=none, fill=none, at={(0.5,1.03)},anchor=north,font=\small},
    legend columns=2,
    legend image code/.code={\draw[##1, draw=none] (0em,-0.2em) rectangle (0.6em,0.4em);},
    axis line style={draw=none},
    axis on top,
    tick style={draw=none},
    yticklabel={~},
    clip=false,
    enlarge y limits=0,
    ,
    nodes near coords always on top/.style={
        every node near coord/.append style={
            anchor=south,
            rotate=0,
            font=\small,
            inner sep=0.2em,
        },
    },
    nodes near coords={
        \pgfmathprintnumber[fixed,fixed zerofill,precision=1]{\pgfplotspointmeta}\%
    },
    nodes near coords always on top,
]
}

\beginCustomBarAxisFirst{left,q,TI\textunderscore QUBIT,a,b,x,c,d,y,right}
\addplot[style={draw=none, fill=\colorQubitLight, mark=none}] coordinates {
    (TI\textunderscore QUBIT, 44.66)
};

\addplot[style={draw=none, pattern color=\colorQubitBB, pattern=north east lines, mark=none}] coordinates {
    (TI\textunderscore QUBIT, 89.85)
};
\end{axis}

\beginCustomBarAxis{left,q,v,u,c,d,BARE\textunderscore QUTRIT,e,f,y,right}
\addplot[style={draw=none, fill=\colorQutritLight, mark=none}] coordinates {
    (BARE\textunderscore QUTRIT, 94.92)
};
\end{axis}

\beginCustomBarAxis{left,q,u,c,d,x,e,f,DRESSED\textunderscore QUTRIT,right}
\addplot[style={draw=none, fill=\colorQutritLight, mark=none}] coordinates {
    (DRESSED\textunderscore QUTRIT, 96.08)
};
\end{axis}

\end{tikzpicture}

%% file: txt/9discussion.tex
\section{Discussion} \label{sec:discussion}
Figure~\ref{fig:simulation_results} demonstrates that our \textcolor{brown}{QUTRIT} construction (orange bars) significantly outperforms the ancilla-free \textcolor{brown}{QUBIT} benchmark (blue bars) in fidelity (success probability) by more than 10,000x.

For the \textcolor{brown}{SC}, \textcolor{brown}{SC+T1}, and \textcolor{brown}{SC+GATES} noise models, our qutrit constructions achieve between 57-83\% mean fidelity, whereas the ancilla-free qubit constructions all have almost 0\% fidelity. Only the lowest-error model, \textcolor{brown}{SC+T1+GATES} achieves modest fidelity of 26\% for the \textcolor{brown}{QUBIT} circuit, but in this regime, the qutrit circuit is close to 100\% fidelity.

The trapped ion noise models achieve similar results--the \linebreak \textcolor{brown}{DRESSED\_QUTRIT} and the \textcolor{brown}{BARE\_QUTRIT} achieve approximately 95\% fidelity via the \textcolor{brown}{QUTRIT} circuit, whereas the \textcolor{brown}{TI\_QUBIT} noise model has only 45\% fidelity. Between the dressed and bare qutrits, the dressed qutrit exhibits higher fidelity than the bare qutrit, as expected. Moreover, as discussed in Appendix~\ref{subsec:trapped_ion_appendix}, the dressed qutrit is resilient to leakage errors, so the simulation results should be viewed as a lower bound on its advantage over the qubit and bare qutrit.

Based on these results, trapped ion qutrits are a particularly strong match to our qutrit circuits. In addition to attaining the highest fidelities, trapped ions generally have all-to-all connectivity \cite{brown2016co} within each ion chain, which is critical as our circuit construction requires operations between distant qutrits.

The superconducting noise models also achieve good fidelities. They exhibit a particularly large advantage over ancilla-free qubit constructions because idle errors are significant for superconducting systems, and our qutrit construction significantly reduces idling (circuit depth). However, most superconducting quantum systems only feature nearest-neighbor or short-range connectivity. Accounting for data movement on a nearest-neighbor-connectivity 2D architecture would expand the qutrit circuit depth from $\log{N}$ to $\sqrt{N}$ (since the distance between any two qutrits would scale as $\sqrt{N}$). However, recent work has experimentally demonstrated fully-connected superconducting quantum systems via random access memory \cite{Quantum_RAM}. Such systems would also be well matched to our circuit construction.

For completeness, Figure~\ref{fig:simulation_results} also shows fidelities for the \linebreak \textcolor{brown}{QUBIT+ANCILLA} circuit benchmark, which augments the ancilla-free \textcolor{brown}{QUBIT} circuit with a single dirty ancilla. Since \textcolor{brown}{QUBIT+ANCILLA} has linearity constants $\sim 10$x better than the ancilla-free qubit circuit, it exhibits significantly better fidelities. While our \textcolor{brown}{QUTRIT} circuit still outperforms the \textcolor{brown}{QUBIT+ANCILLA} circuit, we expect a crossing point where augmenting a qubit-only Generalized Toffoli with enough ancilla would eventually outperform \textcolor{brown}{QUTRIT}. However, we emphasize that the gap between an ancilla-free and constant-ancilla construction for the Generalized Toffoli is actually a fundamental rather than an incremental gap, because:
\begin{itemize}
    \item Constant-ancilla constructions prevent circuit parallelization. For example, consider the parallel execution of $N/k$ disjoint Generalized Toffoli gates, each of width $k$ for some constant $k$. An ancilla-free Generalized Toffoli would pose no issues, but an ancilla-augmented Generalized Toffoli would require $\Theta(N/k)$ ancilla. Thus, constant-ancilla constructions can impose a choice between serializing to linear depth or regressing to linear ancilla count. The Incrementer circuit in Figure~\ref{fig:incrementer} is a concrete example of this scenario--any multiply-controlled gate decomposition requiring a single clean ancilla or more than 1 dirty ancilla would contradict the parallelism and reduce runtime.
    \item Even if we only consider serial circuits, given the exponential advantage of certain quantum algorithms, there is a significant practical difference between operating at the ancilla-free frontier and operating just a few data qubits below the frontier.
\end{itemize}

While we only performed simulations up to 14 inputs in width, we would see an even bigger advantage in larger circuits because our construction has asymptotically lower depth and therefore asymptotically lower idle errors. We also expect to see an advantage for the circuits in Section~\ref{sec:application to algorithms} that rely on the Generalized Toffoli, although we did not explicitly simulate these circuits.

Our circuit construction and simulation results point towards promising directions of future work that we highlight below:
\begin{itemize}
    \item A number of useful quantum circuits, especially arithmetic circuits, make extensive use of multiply-controlled gates. However, these circuits are typically pre-compiled into single- and two- qubit gates using one of the decompositions from prior work, usually one that involves ancilla qubits. Revisiting these arithmetic circuits from first principles, with our qutrit circuit as a new tool, could yield novel and improved circuits like our Incrementer circuit in Section \ref{sec:incrementer}.
    \item Relatedly, we see value in a logic synthesis tool that injects qutrit optimizations into qubit circuits, automated in fashion inspired by classical reversible logical synthesis tools \cite{RevKit, ReversibleLogicSynthesis}.
    \item While $d=3$ qutrits were sufficient to achieve the desired asymptotic speedups for our circuits of interest, there may be other circuits that are optimized by qudit information carriers for larger $d$. In particular, we note that increasing $d$ and thereby increasing information compression may be advantageous for hardware with limited connectivity.
\end{itemize}

Independent of these future directions, the results presented in this work are applicable to quantum computing in the near term, on machines that are expected within the next five years. The net result of this work is to extend the frontier of what is computable by quantum hardware, and hence to accelerate the timeline for practical quantum computing, rather than waiting for better hardware. Emphatically, our results are driven by the use of qutrits for \textit{asymptotically} faster ancilla-free circuits. Moreover, we also improve linearity constants by two orders of magnitudes. Finally, as verified by our open-source circuit simulator coupled with realistic noise models, our circuits are more reliable than qubit-only equivalents. Our results justify the use of qutrits as a path towards scaling quantum computers.

%% file: txt/detailed_noise_model.tex
\section{Detailed Noise Model}
\label{app:detailed_noise_models}

We chose noise models that represent realistic near-term machines. We first present a generic, parametrized noise model in that is roughly applicable to all quantum systems. Next, we present specific parameters, under the generic noise model, that apply to near-term superconducting quantum computers. Finally, we
 present a specific noise model for \textsuperscript{171}Yb\textsuperscript{+}  trapped ions.

\subsection{Generic Noise Model}
The general form of a quantum noise model is expressed by the Kraus Operator formalism which specifies a set of matrices, $\{K_i\}$, each capturing an error channel. Under this formalism, the evolution of a system with initial state $\sigma = \ket{\Psi}{\bra{\Psi}}$ is expressed as a function $\mathcal{E}(\sigma)$, where:
\begin{align}
    \mathcal{E}\left(\sigma\right) = \mathcal{E}\left(\ket{\Psi}\bra{\Psi}\right) = \sum_i K_i \sigma K_i^\dagger
\end{align}
where $\dagger$ denotes the matrix conjugate-transpose.

\subsubsection{Gate Errors}
For a single qubit, there are four possible error channels: no-error, bit flip, phase flip, and phase+bit flip. These channels can be expressed as products of the Pauli matrices:
$$X = \begin{pmatrix}
0 & 1 \\
1 & 0
\end{pmatrix} \text{\quad and \quad} Z = \begin{pmatrix}
1 & 0 \\
0 & -1
\end{pmatrix}$$
which correspond to bit and phase flips respectively. The no-error channel is $X^0 Z^0 = I$ and the phase+bit flip channel is the product $X^1 Z^1$.

In the Kraus operator formalism, we express this single-qubit gate error model as
\begin{align}
    \mathcal{E}(\sigma) = \sum_{j=0}^{1} \sum_{k=0}^{1} p_{jk} (X^j Z^k) \sigma (X^j Z^k)^{\dagger}
\end{align}
where $p_{jk}$ denotes the probability of the corresponding Kraus operator.

This gate error model is called the Pauli or depolarizing channel. We assume all error terms have equal probabilities, i.e. $p_{jk} = p_1$ for $j,k \neq 0$. This assumption of symmetric depolarizing is standard and is used by most noise simulators \cite{QX_Simulator}. Under this model, the error channel simplifies to:
\begin{align} \mathcal{E}(\sigma) = (1-3p_1)\sigma + \sum_{jk \in \{0,1\}^2 \setminus 00} p_1(X^j Z^k) \sigma (X^j Z^k)^{\dagger} \end{align}

For two-qubit gate errors, the Kraus operators are the Cartesian product of the two single-qubit gate error Kraus operators, leading to the noise channel:
\begin{align}
    \mathcal{E}(\sigma) = (1-15p_2)\sigma + \sum_{jklm \in \{0,1\}^4 \setminus 0000} p_2 K_{jklm} \sigma K_{jklm}^{\dagger}
\label{eq:two qubit gate error}
\end{align}
where $p_2$ is the probability of each error term and $K_{jklm}=X^j Z^k \otimes X^l Z^m$.

Next, for qutrits, we have a similar form, except that there are now more possible error channels. We now use the generalized Pauli matrices:
$$X_{+1} = \begin{pmatrix}
0 & 0 & 1 \\
1 & 0 & 0 \\
0 & 1 & 0
\end{pmatrix}
\text{\quad and \quad}
Z_3 = \begin{pmatrix}
1 & 0 & 0 \\
0 & e^{2 \pi i / 3} & 0 \\
0 & 0 & e^{4 \pi i /3}
\end{pmatrix}$$
The Cartesian product of $\{I, X_{+1}, X_{+1}^2\}$ and $\{I, Z_3, Z_3^2\}$ constitutes a basis for all 3x3 matrices. Hence, this Cartesian product also constitutes the Kraus operators for the single-qutrit gate error \cite{QutritErrorChannels, Grassl, Miller}:
\begin{align}
    \mathcal{E}(\sigma) = (1-8p_1)\sigma + \sum_{jk \in \{0,1,2\}^2 \setminus 00} p_1(X_{+1}^j Z_{3}^k) \sigma (X_{+1}^j Z_{3}^k)^{\dagger}
\end{align}
and similarly, the two-qutrit gate error channel is:

\begin{align}
    \mathcal{E}(\sigma) = (1-80p_2)\sigma + \sum_{\substack{jklm \in \\ \{0,1,2\}^4 \setminus 0000}} p_2 K_{jklm} \sigma K_{jklm}^{\dagger}
\label{eq:two qutrit gate error}
\end{align}

Note that in this model, the dominant effect of using qutrits instead of qubits is that the no-error probability for two-operand gates diminishes from $1-15p_2$ to $1-80p_2$, as expressed by equations \ref{eq:two qubit gate error} and \ref{eq:two qutrit gate error} respectively.

\subsubsection{Idle Errors}
\label{ref:idle errors}
For qubits, the Kraus operators for amplitude damping are:
\begin{align} K_0 = \begin{pmatrix} 1 & 0 \\ 0 & \sqrt{1 - \lambda_1} \end{pmatrix} \text{\quad and \quad} K_1 = \begin{pmatrix} 0 & \sqrt{\lambda_1} \\ 0 & 0 \end{pmatrix}
\end{align}

For qutrits, the Kraus operator for amplitude damping can be modeled as \cite{Grassl, GhoshAmplitudeDamping}:

$$
K_0 = \begin{pmatrix} 1 & 0 & 0 \\ 0 & \sqrt{1-\lambda_1} & 0 \\ 0 & 0 & \sqrt{1 - \lambda_2} \end{pmatrix}
\text{, }
K_1 = \begin{pmatrix} 0 & \sqrt{\lambda_1} & 0 \\ 0 & 0 & 0 \\ 0 & 0 & 0 \end{pmatrix}
\text{,}
$$
\begin{align}
\text{ and }
K_2 = \begin{pmatrix} 0 & 0 & \sqrt{\lambda_2} \\ 0 & 0 & 0 \\ 0 & 0 & 0 \end{pmatrix}
\end{align}

As discussed in Section~\ref{subsec:noise simulation}, these noise channels are incoherent (non-unitary), which means that the probability of each error occurring depends on the current state. Specifically, the probability of the $K_i$ channel affecting the state $\ket{\Psi}$ is $\|K_i\ket{\psi}\|^2$ \cite{Nielsen}.

\subsection{Superconducting QC}
We picked four noise models based on superconducting quantum computers that are expected in the next few years. These noise models comply with the generic noise model above and are thus parametrized by $p_1$, $p_2$, $\lambda_1$, and $\lambda_2$. The $\lambda_m$ terms are given by \cite{GhoshAmplitudeDamping}:
\begin{align}
    \lambda_m = 1 - e^{-m \Delta t / T_1}
    \label{eq:lambda_m}
\end{align}
where $\Delta t$ is the duration of the idling and $T_1$ is associated with the lifetime of each qubit.

\subsection{Trapped Ion $^{171}$Yb$^+$ QC}
\label{subsec:trapped_ion_appendix}
Based on calculations from experimental parameters for the trapped ion qutrit, we know the specific Kraus operator types for the error terms, which deviate slightly from those in the generic error model. The specific Kraus operator matrices are provided at our GitHub repository \cite{QutritsGithub}. 

\begin{sloppypar}
We chose three noise models: \textcolor{brown}{TI\_QUBIT}, \textcolor{brown}{BARE\_QUTRIT}, and \textcolor{brown}{DRESSED\_QUTRIT}.  Both
\textcolor{brown}{TI\_QUBIT} and \textcolor{brown}{DRESSED\_QUTRIT} take advantage of clock states and thus have very small idle errors. They both would be ideal candidates for a qudit. The \textcolor{brown}{BARE\_QUTRIT} will suffer more from idle errors as it is not strictly defined on clock states but will require less experimental resources to prepare. Idle errors are very small in magnitude and manifest as coherent phase errors rather than amplitude damping errors as modeled in Section~\ref{subsubsec:idle errors}. We also do not consider leakage errors.  These errors could be handled for Yb\textsuperscript{+} by treating each ion as a $d=4$ qudit, regardless of whether we use it as a qubit or a qutrit.
\end{sloppypar}